\documentclass{cpbtex}
\usepackage{CJK}
\usepackage[format=default,indent=0pt,labelfont=bf,justification=justified,singlelinecheck=false]{caption}
\usepackage{longtable}
\usepackage{graphicx} 
\usepackage{datetime}

\usepackage{xcolor}
\definecolor{lightgray}{RGB}{200, 200, 200} 
\renewcommand{\textbf}[1]{#1} 

\usepackage{cite} 
\let\oldcite=\cite
\renewcommand{\cite}[1]{\textsuperscript{\oldcite{#1}}}
\usepackage{comment}
\usepackage{url}
\usepackage{breakurl}
\usepackage{multicol} 
\usepackage{lineno} 
\usepackage{bm} 

\begin{document}
\begin{CJK*}{GBK}{song}

\title{A Compact X-Ray Laser with Ion Source and Crystal Cavity
\thanks{Project supported by the Research Foundation for Higher Level Talents of West Anhui University (Grant No. WGKQ2021005).}}



\author{\ Shuang Li$^{1,2,3}$
\thanks{Corresponding author. E-mail: shuangli09@fudan.edu.cn, or lishuangwuli@126.com}\\	
$^{1}${School of Electrical and Optoelectronic Engineering, West Anhui University,}\\{ Luan 237012, China.}\\
$^{2}${Institute of Applied Physics and Computational Mathematics,}\\{Beijing 100088, China}\\ 
$^{3}${Shanghai EBIT Lab, Institute of Modern Physics, Department of Nuclear} \\{Science and Technology, Fudan University, Shanghai 200433, China}
}


\maketitle

\begin{abstract}
\noindent 

X-ray free-electron lasers (XFELs) are renowned for their high brightness, significantly impacting biology, chemistry, and nonlinear X-ray optics. However, current XFELs are large, expensive, and exhibit significant shot-to-shot instability. Here, we propose a novel compact apparatus for generating X-ray lasers. The setup integrates an ion source to produce highly charged ions as the gain medium. The X-ray optical cavity employs crystal Bragg diffraction for high reflectivity at large angles, and two parabolic compound refractive lenses (CRLs) focus the X-rays. Pumping is achieved through electron collision excitations.
This X-ray laser offers compact dimensions, reduced costs, and enhanced coherence, positioning it as a promising seed for XFELs. With further optimization, this device has the potential to rival XFELs and revolutionize both scientific research and industrial applications.

\end{abstract}


\section{Introduction}
\label{sec.intro}

High-intensity, high-brightness X-ray lasers have expanded the horizons of various fields, including structural biology\cite{Neutze24N626p,Chatzimagas23PRL131p,Wang21N595p516}, nanometer-scale\cite{Cho21S373p1068} and organic molecular crystal\cite{Takaba23NC15p491} structures, the dynamic processes of atomic and molecular systems\cite{Wang21N595p516,Giannessi21N595p496}, nuclear control\cite{Cho21S373p1068}, strong-field physics\cite{Li22PRL129p}, nonlinear x-ray optics\cite{Wang24PRL132p,Li22PRL129p}, single-particle imaging\cite{Wang24PRL132p}, nanolithography\cite{Fu20CP3p}, ultrafast dynamics studies\cite{Fu20CP3p}, nuclear clocks\cite{Shvydko23Np}, ultra-high-precision spectroscopy\cite{Shvydko23Np}, and extreme metrology in the high-energy X-ray regime\cite{Shvydko23Np}.

Currently, the mechanisms for achieving x-ray amplification primarily include those based on plasmas \cite{Daido02ROPIP65p1513, Namba22A10p128, Lyu20SR10p} and high-order harmonic generation (HHG) \cite{Fu20CP3p}. However, these methods are not widely adopted due to their relatively low energy yield \cite{Fu20CP3p, Reagan14PRA89p}. Additionally, HHG requires significant improvements and research and development toward shorter wavelengths (less than 30 nm) \cite{Pellegrini16ROMP88p}. It is unlikely that wavelengths below 10 nm from HHG will become feasible in the next couple of years \cite{Pellegrini16ROMP88p}.	
In contrast, the mainstream technology currently involves relativistic electron-based free-electron x-ray lasers \cite{Pellegrini16PST169p, Huang23NSAT34p}, which offer superior performance for x-ray amplification.

X-ray free-electron lasers (XFELs) are characterized by their high brightness. For example, their peak intensity is ten orders of magnitude higher than the brightest conventional X-ray sources \cite{Yabashi17NP11p12}. High-speed relativistic electrons are essential for XFEL operation \cite{Pellegrini16PST169p}. However, the acceleration capabilities of conventional accelerators necessitate very long devices \cite{Giannessi21N595p496}, typically ranging from 0.7 to 3 kilometers \cite{Prat20NP14p748}, with costs on the order of 1 billion US dollars. For instance, the XFEL at the SHINE facility in Shanghai has a total length of about 3.1 kilometers\cite{Huang23NSAT34p}, with an estimated cost of 1.5 billion US dollars (see \url{https://accelconf.web.cern.ch/ipac2024/pdf/TUZN1.pdf}).
The large size\cite{Wang21N595p516}, high construction costs\cite{Wang21N595p516}, and substantial operational expenses severely limit the application ranges of XFELs \cite{Lv22PRL128p,Nugent10AiP59p1}.

To reduce the size of such equipment, researchers have explored replacing conventional acceleration with plasma-based or laser-based methods \cite{Wang21N595p516}. Ultra-high accelerating gradients (exceeding 100 GV m$^{-1}$) \cite{Wang21N595p516} from the two methods offer the potential to reduce the equipment size by more than an order of magnitude \cite{Labat23NP17p150}. 
However, the lack of reliability and reproducibility of the electron beams remains the primary issue hindering further application of the methods \cite{Wang21N595p516}.

Present-day X-ray Free Electron Lasers (XFELs) are primarily based on self-amplified spontaneous emission (SASE). SASE XFELs can generate pulses with a high degree of transverse coherence \cite{Labat23NP17p150}. However, due to their noise-based starting point \cite{Rohringer12N481p488, Labat23NP17p150}, they suffer from poor shot-to-shot stability \cite{Allaria23NP17p841} and low longitudinal coherence \cite{Labat23NP17p150, Allaria23NP17p841}, typically exhibiting spiky temporal and spectral distributions \cite{Labat23NP17p150, Wang21N595p516, Rohringer12N481p488, Allaria23NP17p841}.	
To address these limitations, external seeding was proposed \cite{Labat23NP17p150}. External seeding can significantly improve both shot-to-shot stability and longitudinal coherence by providing a coherent seed signal. For hard X-ray wavelengths, however, the lack of suitable seed lasers has necessitated the use of self-seeding techniques. Self-seeding was successfully implemented at LCLS in 2012 \cite{Bostedt16RoMP88p15007}. Despite this achievement, self-seeded XFELs still exhibit poor shot-to-shot stability\cite{Huang21I2p, Allaria23NP17p841}. Additionally, the self-seeding unit is complex and cumbersome\cite{Lv22PRL128p}, which adds to the operational challenges.

To further improve the coherence, an X-ray free-electron laser (XFEL) based on an X-ray optical cavity has been proposed \cite{Kim08PRL100p}. Experimental evidence has demonstrated that diamond can achieve nearly 100\% Bragg diffraction efficiency \cite{Shvydko11NP5p539} and can survive the power load within the cavity while maintaining very high reflectivity \cite{Kolodziej18JoSR25p1022}. These properties make diamond an ideal X-ray reflector for the cavity.	
The direct observation of sustained stable X-ray circulation provides the most direct evidence to date that cavity-based X-ray free-electron lasers and other cavity-based hard X-ray systems are feasible \cite{Margraf23NP17p878, Allaria23NP17p841}. Currently, the SHINE (Shanghai High Repetition Rate XFEL and Extreme Light Facility) XFEL facility is considering implementing an XFEL cavity \cite{Huang23NSAT34p}. Meanwhile, teams at Spring-8 \cite{Cho21S373p1068}, SLAC \cite{Allaria23NP17p841, Cho21S373p1068}, and the European XFEL \cite{Allaria23NP17p841} are actively working on building test cavities.

Here, we propose an alternative approach for compact X-ray lasers that utilizes an X-ray cavity based on an ion source. The high reflectivity X-ray cavity achieved through Bragg diffraction of crystals. This approach offers several significant advantages over free-electron X-ray lasers (FELs):	1. Compact Device Size: The new design results in a smaller and more compact apparatus, which is easier to install and operate.
2. Reduced Construction Costs: The simplified design and fewer components lead to lower overall construction costs.
3. Lower Maintenance Expenses: The compact and simplified system requires less expensive maintenance.	These benefits make it feasible to conduct a large number of X-ray laser experiments and facilitate widespread industrial applications.

\section{Components of the Device}
\label{sec.dev}

\setcounter{figure}{0}
\begin{figure}[ht!]
	\centering
	\includegraphics[height=2.0in]{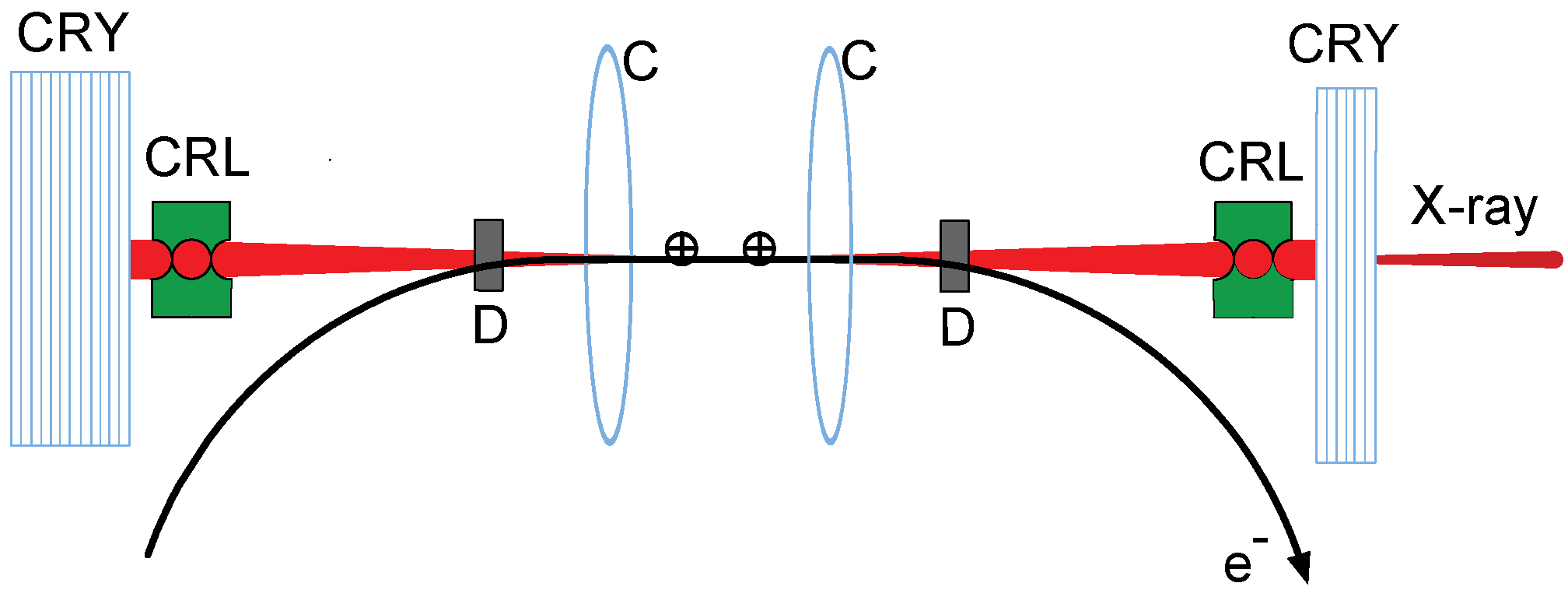}
	\caption{ \label{fig1} Scheme of cavity-based X-ray laser utilizing ion source. 
	The X-ray optical cavity is composed of two high-reflectivity flat crystals (CRYs) and two parabolic compound refractive lenses (CRLs). The right CRY is designed to have specific transmittance, enabling the desired X-ray output. The two CRLs focus the beam and, along with the flat CRYs, create a stable optical cavity.	
	Highly charged ions in excited states, generated in a special ion source, are unstable. One of the spontaneously emitted photons from these ions travels along the horizontal optical axis, acting as seed light. This seed light oscillates within the optical cavity, gaining amplification.
	As for the special ion source, an electron gun (not illustrated) emits an electron beam, which enters the left deflector (D) from the lower left of the figure along the solid black line. After deflection, the electron beam enters the strong magnetic field generated by two Helmholtz coils(C, or a solenoid). In the magnetic field, the electron beam is compressed and collides with the injected gas or metal vapor atoms, gradually ionizing them to form highly charged ions ($\oplus$). The ions serve as the gain medium of the laser, and under the continued collisions from the electron beam, they may gain energy and stay as an excited state, even potentially exceeding the population threshold. The electron beam then passes through the right deflector (D) and is deflected out from the lower right of the figure, ultimately being collected by a collector (not illustrated). Finally, the entire system (including the X-ray cavity\cite{Margraf23NP17p878} ) is enclosed in a vacuum environment (not illustrated).
	}
\end{figure}

Figure~\ref{fig1} illustrates a schematic diagram of the newly proposal X-ray laser. The three important components of this device are X-ray optical cavity, gain medium and pump.

The X-ray optical cavity is composed of two high-purity crystals (CRYs) and two Compound Refractive Lens (CRLs). 
Due to the unique properties of X-rays, the high-reflectivity of the cavity can be obtained by using crystals composed of low-Z atoms with high Debye temperature\cite{Kim08PRL100p}, such as C (diamond), BeO, SiC, or $\alpha-$Al$_2$O$_3$ (sapphire)\cite{Kim08PRL100p}.

Both theoretical calculations based on dynamical diffraction theory\cite{Margraf23NP17p878} and experimental studies\cite{Shvydko11NP5p539} confirm that at specific incident X-ray photon energies\cite{Cho21S373p1068}, utilizing the Bragg diffraction of specific crystal orientations\cite{Cho21S373p1068} can achieve high reflectivity at large incident angles, including normal incidence\cite{Shvydko17MB42p437}.
For instance, experimental findings indicate that perfect diamond crystals can attain reflectivity up to 99\%\cite{Shvydko11NP5p539,Margraf23NP17p878}.
In figure~\ref{fig1}, the crystal on the left is designed with high reflectivity, whereas the thin, defect-free, and strain-free drumhead crystal\cite{Shvydko17MB42p437} (less than 0.01 mm\cite{Kim08PRL100p}) on the right, with around 5\% transmission\cite{Shvydko17MB42p437,Kim08PRL100p}, serves as the output coupler for the X-ray laser.
Numerous alternative methods exist for the output coupling of X-ray lasers\cite{Tang23PRL131p}, including the use of X-ray gratings\cite{Liu24JOSR31p751}, splitters, pin-hole diamond mirrors, or diamond mirrors with doping\cite{Tang23PRL131p}.

To enhance the stability\cite{Margraf23NP17p878} of X-rays during their propagation and effectively control their modes\cite{Margraf23NP17p878}, it is necessary to focus the X-rays\cite{Kim08PRL100p,Liu24JOSR31p751}. Since direct focusing through even a very gentle bending of crystals would significantly reduce the crystal's reflectivity\cite{Kim08PRL100p}, researchers typically use two CRL mirrors made from low-Z materials for focusing. A promising option is to use two parabolic compound refractive lenses (CRLs) made from low-Z materials, such as beryllium\cite{Kim08PRL100p,Liu24JOSR31p751}, for this purpose\cite{Kim08PRL100p}.
And the entire cavity is enclosed in vacuum\cite{Margraf23NP17p878}.

Since crystals with a very narrow reflection bandwidth, typically meV-narrow for X-rays\cite{Shvydko17MB42p437,Kolodziej18JoSR25p1022}, and X-ray lasers emitted by highly charged ions\cite{Daido02ROPIP65p1513,Namba22A10p128,Lyu20SR10p}, which we will detail in the gain medium section, have specific wavelengths, there may be a mismatch. The Doppler broadening caused by the thermal motion of highly charged ions in the ion source can be utilized for adjustment and matching. Additionally, a cavity with a bow-tie structure can tune the X-ray photon energy within a 5\% range\cite{Huang23NSAT34p,Dai12PRL108p}, which should be very helpful.

It is important to note that Figure~\ref{fig1} is only a schematic illustration. In fact, for X-ray lasers of different energies, besides the normal incidence of X-rays mentioned above, other large-angle incidence methods can also be employed. For example, four crystals can be used as reflectors at a 45-degree angle of incidence\cite{Margraf23NP17p878}, or a bow-tie cavity can be adopted\cite{Huang23NSAT34p}.

The second is the gain medium for X-ray lasers, specifically highly charged ions\cite{Daido02ROPIP65p1513,Namba22A10p128,Lyu20SR10p}.
Highly charged ions possess the capability to emit X-rays. Within isoelectronic sequences, where ions share the same number of electrons but differ in nuclear charge ($Z$), a distinct trend emerges: as $Z$ increases, the wavelength of a specific electronic transition decreases, corresponding to an elevation in X-ray energy. To illustrate, consider Li-like ions, characterized by having three electrons. In these ions, aluminum (Al) exhibits a 3d-4f laser transition with a wavelength of 15.5 nm \cite{Namba22A10p128}. By employing the GRASP2K \cite{Jonsson23A11p} and FAC \cite{Gu08CJOP86p675} computational codes, we have determined the wavelengths for iron (Fe) and germanium (Ge) to be 3.2 nm \cite{Li23CPB32p} and 2.1 nm \cite{Li24CPB33p}, respectively. Additionally, our predictive calculations indicate that for an ion with $Z$=92, the wavelength of this transition can be reduced to a remarkable 0.2 nm (forthcoming publication).

Here highly charged ions are produced by a specialized ion source \cite{Donets98RoSI69p614,Currell05ITOPS33p1763,Zschornacka14apap}. 
The ion source is unique as it includes two additional deflectors (D) compared to traditional ones. These deflectors are designed to ensure that the electron beam path only partially overlaps with the X-ray path, with highly charged ions being produced in the overlapping region as the laser working medium.

As illustrated in Figure \ref{fig1}, the electron beam is emitted from an electron gun (not illustrated) located in a zero or low magnetic field area and accelerated into the left deflector D from the lower left of the figure. After deflection, the beam enters three drift tubes composed of three successive cylindrical electrodes and is compressed by a strong magnetic field. This magnetic field is generated by a solenoid or a pair of superconducting Helmholtz coils(C) surrounding the drift tubes. The quasi-monochromatic, high-energy, high-current-density electron beam then collides and sequentially ionizes injected low-charge-state ions or atoms, gradually forming highly charged ions. The radial trapping of the ions is achieved through the space charge of the electron beam, while axial trapping is accomplished by applying a negative bias to the central drift tube relative to the two end drift tubes. After passing through the drift tubes, the electron beam is deflected by the right deflector (D), exits from the lower right of the figure, and is decelerated and collected by an electron collector (not illustrated).
To achieve a sufficiently high ionization rate, the electron beam is electrostatically accelerated to the required energy, typically between 500 eV and 200 keV~\cite{Currell05ITOPS33p1763}. It is important to note that the entire electron beam system is maintained in an ultra-high vacuum environment~\cite{Currell05ITOPS33p1763,Ovsyannikov10JoI5p11002}, better than $1\times10^{-9}$Torr.~\cite{Pikin10JoI5p9003}.

The total number of ions that can be stored in an EBIS is constrained by the electrical capacity of the ion source~\cite{Donets98RoSI69p614,Zschornacka14apap}, denoted as $C_{\text{e}}$.
To estimate $C_{\text{e}}$, we consider a homogeneous electron beam traversing an ion source with trap length $L$, characterized by an electron beam current $I_{\text{e}}$ and electron energy $E_{\text{e}}$. The number of negative charges present in the volume of the beam serves as an indicator of the capacity $C_{\text{e}}$~\cite{Donets98RoSI69p614,Zschornacka14apap}:
\begin{equation}
	\label{eq:ce}
	C_{\text{e}}=1.05\times10^{13} \frac{I_{\text{e}}(\text{A}) L(\text{m})}{\sqrt{E_{\text{e}}(\text{eV})}}
\end{equation}

Given that the electron energy \(E_{\text{e}}\) is determined by a sufficiently high ionization rate, enhancing the trap capacity \(C_{\text{e}}\) requires increasing the current \(I_{\text{e}}\) and the length \(L\). To increase the current \(I_{\text{e}}\), one method is to boost the current intensity of a single electron gun.
For example, using the long-life IrCe cathode, the RHIC EBIS electron gun can achieve currents of 10 A or even higher~\cite{Pikin10JoI5p9003}.
Another possible method we suggest here involves using multiple electron guns in parallel and combining the electron beams.
Additionally, the trap length \(L\) of the ion source can be increased.
For example, to achieve an ion source capacity twice that of the Test EBIS, the primary goal of constructing the RHIC EBIS was to double the ion trap length~\cite{Pikin10JoI5p9003}.   
Although instability issues due to lengthening need to be considered~\cite{Watanabe97JOTPSOJ66p3795}. An alternative approach we suggest here to increasing the length is to place many ion sources in series within the cavity.
In fact, through careful design, the trap capacity \(C_{\text{e}}\) can indeed be increased~\cite{Becker10RoSI81p}. For instance, compared to the Daston EBIS-SC device, which has a trap capacity of 2$\times10^{10}$, the RHIC-EBIS boasts a capacity of up to 2$\times10^{12}$~\cite{Becker10RoSI81p}.

The third is pump. 
The electron beam mentioned above can serve as a pump, they can not only be used to generate highly charged ions but also further collide with these ions, transferring some energy to them. This can excite the ions, may leading to population inversion and even potentially exceeding the inversion threshold.
In addition to the direct collision excitation mechanism, particle inversion can also be achieved through recombination of highly charged ions~\cite{Namba22A10p128}.

Of course, in addition to these three main components, the entire laser system has other auxiliary parts, such as systems for handling the large amounts of waste heat.


\section{Conclusion}
\label{sec.conc}
In summary, we propose an innovative compact X-ray laser scheme based on highly charged ions. Compared to the existing mainstream free-electron X-ray lasers, this scheme offers better coherence, smaller size, and lower cost. We anticipate that this scheme could serve as a seed source for free-electron X-ray lasers and, with increased capacity of newly designed  specialized ion sources, potentially totally replace free-electron X-ray lasers at specific wavelengths.

\section*{Acknowledgment}
This research is supported by the Research Foundation for Higher Level Talents of West Anhui University (Grant No. WGKQ2021005).
The author is grateful to Yan Wang, Yuxuan Li, Zhengshuang Liu, Chen Zhao, and Hao Zhou for their contributions in preparing this manuscript.

\bibliographystyle{unsrt}
\bibliography{ref-xray}

\end{CJK*}  
\end{document}